\RequirePackage{lineno}
\documentclass[12pt,prl,preprint,letterpaper,linenumbers]{article}
\usepackage{lineno}
\usepackage[english]{babel}
\usepackage[utf8]{inputenc}
\usepackage{amsmath}
\usepackage{graphicx}
\usepackage{adjustbox}
\usepackage{xspace}
\usepackage{hyperref}
\usepackage{xcolor}
\usepackage{makecell}
\usepackage{authblk}

\newcolumntype{L}{>{\centering\arraybackslash}m{8cm}}

\usepackage[margin=1.0in]{geometry}

\title{Summary of Workshop on Common Neutrino Event Generator Tools
}
\author[1]{Josh Barrow}
\author[2]{Minerba Betancourt}
\author[3]{Linda Cremonesi}
\author[4]{Steve Dytman}
\author[2]{Laura Fields}
\author[5]{Hugh Gallagher}
\author[2]{Steven Gardiner}
\author[2]{Walter Giele}
\author[2]{Robert Hatcher}
\author[2]{Joshua Isaacson}
\author[6]{Teppei Katori} 
\author[2]{Pedro Machado}
\author[7]{Kendall Mahn}
\author[8]{Kevin McFarland}
\author[9]{Vishvas Pandey}
\author[10]{Afroditi Papadopoulou}
\author[11]{Cheryl Patrick}
\author[12]{Gil Paz}
\author[7]{Luke Pickering}
\author[2,13]{Noemi Rocco}
\author[14]{Jan Sobczyk}
\author[5]{Jeremy Wolcott}
\author[8]{Clarence Wret}

\affil[1]{University of Tennessee, Knoxville, TN 37996, USA}
\affil[2]{Fermi National Accelerator Laboratory, Batavia, Illinois 60510, USA}
\affil[3]{Queen Mary University of London, London E1 4NS, United Kingdom}
\affil[4]{ University of Pittsburgh, Pittsburgh, Pennsylvania 15260, USA}
\affil[5]{Tufts University, Medford, Massachusetts 02155, USA}
\affil[6]{King's College London, London WC2R 2LS, UK}
\affil[7]{Michigan State University, East Lansing, Michigan 48824, USA}
\affil[8]{University of Rochester, Rochester, New York 14627 USA}
\affil[9]{ University of Florida, Gainesville, FL 32611, USA}
\affil[10]{Massachusetts Institute of Technology, Cambridge, Massachusetts 02139, USA}
\affil[11]{University College
London, Gower Street, London WC1E 6BT, United Kingdom}
\affil[12]{Wayne State University, Detroit, Michigan 48201, USA}
\affil[13]{Argonne National Lab, Argonne IL 60439, USA}
\affil[14]{Wroclaw University, 50-204 Wroclaw, Poland}

\begin{document}

\maketitle
\section{Introduction }

Neutrino event generators are a critical component of the simulation chains of many current and planned neutrino experiments, including the Deep Underground Neutrino Experiment (DUNE)~\cite{DUNE} and Hyper-Kamiokande (HK)~\cite{Abe:2015zbg}. Simulating neutrino interactions with nuclei is particularly complex for accelerator-based neutrino experiments, which typically use neutrino beams with energy ranging between hundreds of MeV to tens of GeV.  The dominant neutrino-nucleus interaction changes across this energy range, from relatively simple quasi-elastic scattering at the lowest energies, to deep inelastic scattering at the highest energies.  The heavy nuclei that compose modern neutrino detectors introduce additional complications, as effects of the nuclear environment must also be modeled.  

Neutrino-nucleon interactions, are an area of very active theoretical development and are informed by a recent wealth of neutrino scattering measurements from MiniBooNE~\cite{miniboone}, MINERvA~\cite{minerva}, MicroBooNE\cite{microboone}, NOvA\cite{nova} and T2K~\cite{t2k} that illustrate the deficiencies of historical models.  A panoply of theoretical models have become available in recent years (see examples in Table~\ref{table_models}).
Despite this level of attention from the theory community, neutrino
experiments have been slow to adopt the latest model improvements. This is
primarily due to a software development bottleneck: several years have
typically been required for a completed theory calculation to become available
in a neutrino event generator.  This not only prevents experiments from using the latest models but also creates a very slow feedback loop for theorists validating their models against data.  
 
 The delay between model development and availability in generators comes about partially because implementing models into a generator is difficult.  Another factor is that there are several separate generators on the market, and experiments generally do not have the resources to simulate events in all generators.  The available generators include:
 \begin{itemize}
     \item GENIE~\cite{Andreopoulos:2009rq}, developed and maintained by an international collaboration and used as a primary generator by Fermilab experiments including NOvA, MINERvA, SBN~\cite{SBN} and DUNE; used as a secondary generator by other experiments including T2K and HK.
     \item GiBUU~\cite{Buss:2011mx}, notable for its unified neutrino-nucleus and hadron transport modeling; not currently used as the primary generator by experiments but frequently used to compared to primary generators.
     \item NEUT~\cite{Hayato:2009zz}, initially developed for Kamiokande and used as a primary generator by T2K, Super-Kamiokande (SK)~\cite{superk}, and HK.  
     \item NuWro~\cite{Golan:2012rfa}, a theory-oriented generator developed at the University of Wroclaw; not used as the primary generator of any experiment but commonly used to cross check primary generators.  
 \end{itemize}
 Although all of these generators perform similar functions---taking in neutrino energy spectra and detector geometries (or target nuclei) and producing lists of final-state particles---the entire simulation chain is reimplemented separately in each generator and there is little commonality to interfaces.  For example, input and output event formats and configuration files are completely different.  There is also no standard interface for theorists wishing to implement models in generators.   There are significant differences across the generators in ability to handle complex detector geometries and neutrino kinematic (``flux") distributions. 
 
 Validating a neutrino model against data is currently a multi-stage process that typically requires a theorist to deliver a model to a member of a generator development team, who manually adds it to the generator. Generator teams vary in size, but none consist of more than a handful of active developers, all of whom have additional responsibilities. Once a model is implemented in a generator, it can be compared with data.
This may be done by using the generator to
directly compute the relevant differential cross sections (similar to
what a theorist might produce for a paper). Alternatively, the generator
might be incorporated into an experiment's software stack and used as
part of an end-to-end simulation of beam production, particle interactions,
and the detector response (as is done for oscillation analyses).
 
 There is broad recognition in the neutrino community of the need to ease the process of transferring models from theorists to experiments, and that common interfaces across generators are essential to meeting this need.  Common generator interfaces were some of the topics discussed at the Workshop on ``Testing and Improving Models of Neutrino Nucleus Interactions in Generators", held at the ECT* in Trento, Italy in June, 2019.  At that workshop, a series of concrete steps were identified that would substantially improve the ability of neutrino experiments to use the latest neutrino-nucleus interaction models.  
 
 To build on the momentum established at the Trento workshop, Fermilab held a workshop Jan 8-10, 2020 aimed at developing a technical plan for implementing these steps.  In this white paper, we summarize that workshop and the resulting plan, which will allow neutrino experiments to more efficiently use new models, and to compare predictions from different generators. 
 While this plan will benefit many current and future experiments, the work
proposed herein is focused particularly on meeting the anticipated needs of the DUNE and HK experiments.  

\section{Workshop Summary }

The Generator Tools Workshop was held at Fermilab from Jan 8-10, 2020.  The agenda and presentations are available online~\cite{workshop_indico}. The introductory session, which described the goals of the workshop, included
talks on the experience with using multiple generators by LArSoft~\cite{Pordes:2016ycs} and
the potential of using GENIE as a common platform for other generators. Additionally, representatives from each of the four generators' teams were invited to present an overview of their generator and to respond to four prompts:
\begin{itemize}
    \item Give an overview of your flux/geometry driver functionality and input formats
    \item Are there any structural problems that would be needed to solved to use a common flux/geometry driver?
    \item Review the output formats of your generator
    \item Are initial- and final-state interactions modeled at separate stages in your generator?  If not, why not? How difficult, mechanically, would this factorization be?
\end{itemize}
The bulk of the workshop consisted of sessions devoted to discussing a plan to implement common versions of the following generator interfaces:
\begin{itemize}
    \item Flux and geometry driver
    \item Flux and geometry input formats
    \item Generator output file formats
    \item Separation of primary interaction and FSI 
    \item Theory interface 
\end{itemize}
In each workshop session, discussion leaders and note-takers were assigned, and two participants were charged with summarizing the session.  Those summaries are provided in the remainder of this section and the resulting technical plans are discussed in Section~\ref{sec:plan}.  

\subsection{Common Flux and Geometry Driver} 

A flux driver is the portion of a neutrino event generator that consumes information about the neutrino flux and produces neutrino 4-momenta $(E,\vec{p})$ distributed across the surfaces of a detector. Generators often have many different flux drivers to handle different types of neutrino flux inputs. 

A geometry driver is the portion of the generator that takes the neutrino 4-momenta (from the flux driver) and a detector geometry as inputs and produces event vertices distributed across the volume of a detector.  Because it takes into account the distribution of various materials throughout the detector, it requires knowledge of the total interaction cross sections on various nuclei predicted by the generator.  Flux and geometry drivers are frequently integrated together within generators.  

The four currently available generators have implemented these in various ways, which we briefly summarize below.  
\begin{itemize}
    \item GENIE includes a number of flux drivers, capable of consuming BGLRS, FLUKA, and HAKKM atmospheric flux files, fluxes from the JPARC, BNB and NuMI neutrino beam lines, or generalized fluxes provided as histograms or ntuples.  It also employs a geometry driver capable of reading ROOT geometry files, as well as a ``point geometry" driver capable of handling simple mixtures of materials but no spatial information.   
    \item GiBUU does not currently have complete flux or geometry drivers.  Rather, it produces predicted cross sections and final-state kinematic distributions for fixed neutrino energies, and is also able to provide these quantities averaged over a given neutrino energy spectrum, which the user provides as a histogram with a specified format.  Flux distributions for current and recent neutrino experiments are available in GiBUU for user convienence.  Effects of detector geometries and material composition are left to the user.
    \item NEUT includes separate flux/geometry drivers for the T2K near and far detectors.  Both consume neutrino 4-momenta in ntuple format (or histogram format for the far detector).  For the far detector, a fixed beam direction and uniform vertex distribution are assumed, while the near detector driver consumes a detector geometry in ROOT format and the beam direction taken from the flux ntuple.  A simple option capable of consuming a flux histogram and producing events on a single nucleus is also available.  An additional driver for SK atmospheric flux is also available in the SK codebase but is not distributed as part of NEUT.  
    \item NuWro includes a flux driver with six options including specifying a single-flavor beam with a given energy spectrum, a flavor-mixed beam
    with specified spectra, a beam specified by the T2K beam ntuple, and beams with inhomogeneous space and momentum distributions specified by multi-dimensional histograms.  It also has geometry drivers for single isotopes, isotope mixtures, or detector geometries in ROOT format.    
\end{itemize}

In discussions at the workshop, it was clear that significant amounts of person power are being duplicated in the creation and support of these various flux and geometry drivers.  There was significant interest in development of a common driver that can be used by multiple generators.  It was also noted that a common flux/geometry driver would facilitate the creation of ``mini-generators" that would allow theorists to simulate events for a model, using a particular neutrino flux and detector geometry,  without incorporating that model
into the codebase of one of the general-purpose generators.  

\begin{figure} [h] 
\begin{center}
\includegraphics[width=0.7\textwidth]{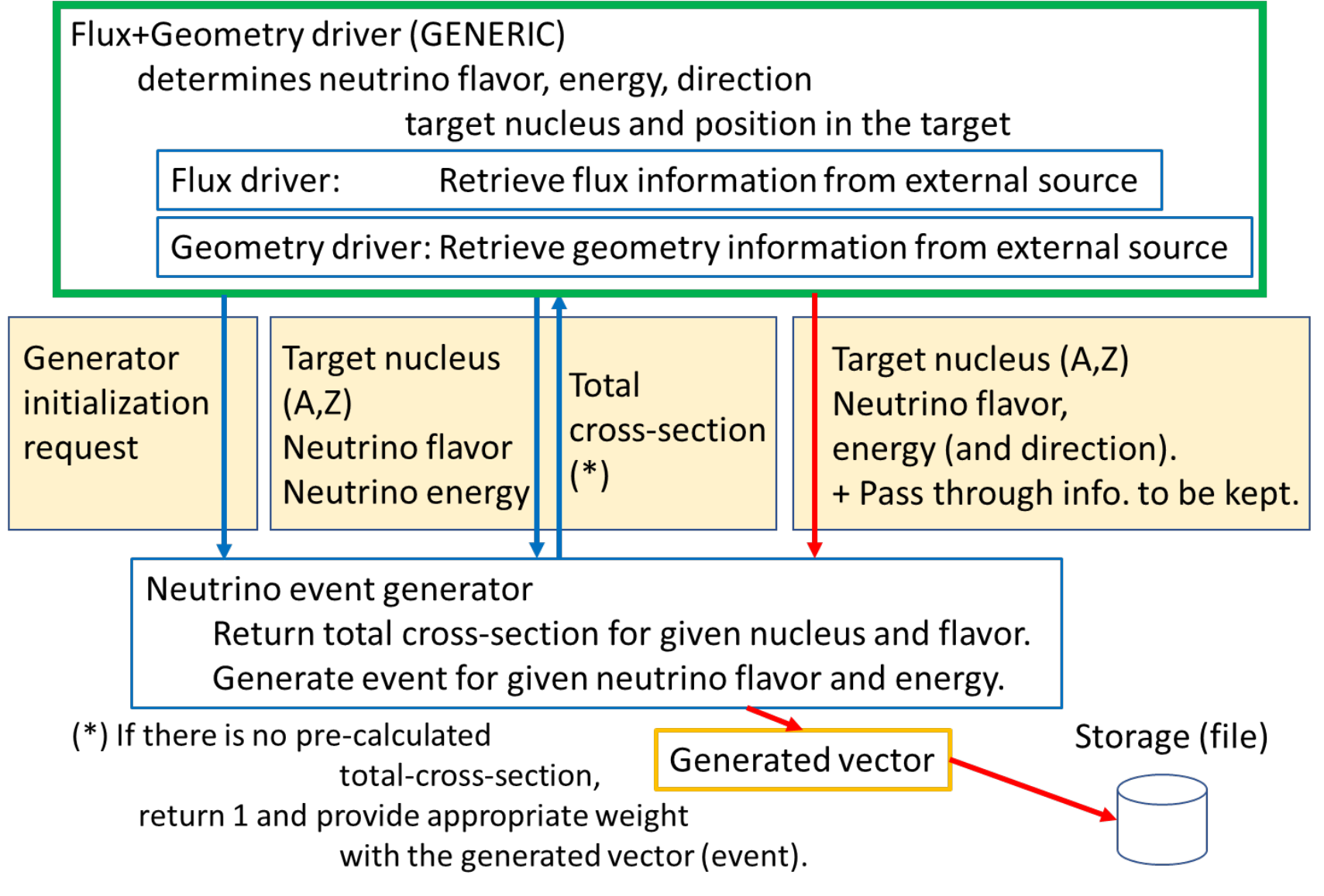}
\end{center}
\caption{\label{fig:hayato_flux}Schematic of a possible interface between generators (blue box) and a community-based flux and geometry driver (green box), created by Yoshinari Hayato.  }
\end{figure}

Workshop participants discussed various options for the creation of a common flux/geometry driver.  In that conversation, two options emerged as leading candidates:  
\begin{itemize}
    \item Extension of the GENIE flux and geometry drivers.  GENIE has the most complete and feature-rich drivers, making them an obvious candidate for adoption by other generators.  The GENIE collaboration has agreed to consider a model wherein GENIE provides hooks to other generators within the GENIE build system, and authors of other generators develop lightweight interfaces between their
software and the GENIE flux/geometry drivers.  This development would be part of the GENIE incubator process, wherein additions are reviewed and validated by the GENIE collaboration before incorporation into GENIE.  An MOU between GENIE and other parties defining responsibilities and licensing would be necessary with this option.  
    \item New community-based flux and geometry drivers.  These would duplicate the functionality already in place in several generators, but would be available for all current and new generators going forward.  This option was viewed favorably by many participants of the workshop, some of whom have already begun writing prototypes.  Because the geometry driver would need to access predicted cross sections from generators, choices will need to be made about the scope and interface of the driver.  One possible choice is outlined in Fig.~\ref{fig:hayato_flux}.  A primary challenge with this option would be to identify the parties responsible for maintenance going forward (e.g. lab, universities, collaborations, or an organization such as NuSTEC).     
\end{itemize}

\subsection{Geometry and Flux Formats for Common Driver} 
\label{sec:flux_summary}
All neutrino generators that currently include a geometry driver take ROOT geometry files as inputs.  Translation tools between other (e.g. Geant4) geometry tools and ROOT exist.  Many experimental simulations use GDML (Geometry Description Markup Language) geometry descriptions that are converted to ROOT.  Therefore, it is clear that any common geometry driver should take ROOT or GDML geometry files as input.  

The situation with flux driver inputs is less straightforward.  Different types of neutrino fluxes (atmospheric, solar, reactor, accelerator-based, etc.) are specified in different ways.  Moreover, even for a single neutrino source, several input formats are possible (histograms of neutrino energy spectra, lists of neutrino 4-vectors, etc).  It is clear that a common flux driver would need to have interfaces to several different types of input formats.  In some cases, it may be possible to adopt a common format for particular sources; e.g. Fermilab maintains an open source library and flux event format called ``dk2nu" (decay to nu) that is used by Fermilab-based neutrino experiments, but could be expanded to other beamlines.  For the moment, it appears simpler to have the common flux driver accept many different types of input.    

\subsection{Output Format}

Generators preserve information such as the initial state nucleon kinematics, details of the hard scatter prior to final-state interactions, and details of final state interactions in output files that are currently different for each generator.  Several collaborations (e.g. NUISANCE\cite{nuisance}, T2K, MINERvA, NOvA) have developed solutions for dealing with these various formats, often in the form of file translators.  An accord between the generators to use a common format would have several uses:
\begin{itemize}
    \item Making the sharing of experiments' custom model adjustments and uncertainties more straightforward (easing the path for comparison and combination of experimental results such as the ongoing effort between NOvA and T2K). 
    \item Facilitating the hand-offs of events between generators (including  mini-generators developed by theorists), allowing one generator to e.g. generate events using a particular hard-scattering model and pass them to another generator for the FSI step. \item Making the process more straightforward when linking external packages (e.g. LArSoft) to several external generators.
    \end{itemize}
    The term 'event format' can have two possible definitions: an in-memory representation used during generation of events, or a representation of the event that is written as output.  For the purposes of this document, we are primarily discussing the event format that the generator passes to external software.   

Several elements required from a common format were identified:
\begin{itemize}
    \item Event-wise
    \begin{itemize}
        \item Snapshots of event particles at various stages (initial state, after primary neutrino-nucleus interaction, and at various steps in the FSI simulation) in the lab frame, parentage, status codes
        \item Primary neutrino-nucleus interaction information such as reaction type and sub-mode if applicable
        \item The interaction vertex location within the nucleus
\item  Information about decays and formation of hadrons from quarks (hadronization).  
\item Neutrino hadron ancestry information (flux)
\item  Generator-specific information describing the interaction process, which allows events to be reweighted as a function
of model parameters. This is a standard technique used by
neutrino experiments to assess  systematic uncertainties associated with model parameters.
    \end{itemize}
    \item Metadata such as generator-specific model configurations and beam configurations 
    \end{itemize}
Many participants at the workshop supported investigation of tools developed by the wider HEP community, such as HepMC.  It was also agreed that the format should support non-neutrino probes (such as electrons, pions, and protons) by design, and that further thought should be put into the question of whether that adds new requirements.  

Several potential complications to development of a common output format were discussed, including:
\begin{itemize}
    \item Different generators delineate primary processes differently.  For example, each generator has different criteria for classifying an event as Deep Inelastic Scattering.  
\item Different primary processes support different options which should be preserved in the event record.  For example, when generating resonant events, the number of allowed nucleon resonances that may be simulated differently by different generators.

\item Choices of working frame (e.g. lab frame or struck nucleon rest frame) may differ
between generators.  For a common format, variables must be stored carefully and clearly.  For a subset of critical
outputs, there will need to be an agreed upon a standard.  This necessitates a community-supported “glossary” of what every possible
entry in the record means as part of the common format.  
\item Nontrivial interactions between model stages will need to be recorded, e.g., at the boundaries between
hadronization, FSI, and unstable particle decays.   Additionally, events that were rejected internally between stages must be appropriately handled.  
\end{itemize}

Two possible paths forward towards a common event record were discussed.  The first would be to create translators.  The NUISANCE collaboration has already developed a basic translator that consumes events from all four generators to produce its internal format.  As the NUISANCE framework does not write its native format as output, that step would require new development.  

The second path would be for generators to write directly to a commonly-defined format.  While not technically complex, this option does require generator buy-in.  This option is more robust in the long run, especially if unit tests, etc. are included directly.  Generators could reasonably write both a native and a common format, as storage for this stage of neutrino simulations is not typically a driver for computing needs. This closely resembles the adoption of HepMC in collider physics.

\subsection{Separation of Primary Interaction and Final-State Interactions (FSI)}  

This portion of the workshop considered the benefits and challenges associated with supporting FSI as a second stage from a generator, separate from its primary interaction simulation(s).  In combination with a common event record that could be used by generators to pass hard-scattering data to the FSI step, this allows cross-fertilization of FSI models and comparison of different primary interaction and FSI models.  It could also accelerate the adoption of best practices and the inclusion of models into different generators.  Use cases for this functionality include:
\begin{itemize}
    \item Understanding which parts of FSI models correlate with observable features in data.  
    \item Quantifying the consequences of violating internal consistency of nuclear models (see further discussion later in this section)
    \item Quickly studying a  mini-generator hard-scattering model with plausible hadron transport.
\end{itemize}

There are a number of pitfalls associated with separation of the primary interaction and FSI.  It inherently supports inconsistency in nuclear models.  The nuclear model (potential, density, etc) used for the primary interaction can be inconsistent with the nuclear model
assumed in the hadron propagation.  For instance, pion production and absorption happen via the same diagram, so it is unphysical to use different models in production and hadron transport.  Separation of the primary interaction and FSI also violates time reversal symmetry, which is clearly incorrect, but the consequences are difficult to quantify without implementing the separation.    

A second pitfall would be associated with workflow maintenance.    The hard-scattering and FSI calculation modules are included in individual generators, but a workflow might involve mixing and matching between generators, so a key question is who maintains documentation and any necessary code to ensure a seamless connectivity between generators.  

A third pitfall involves the coupling of decay and hadronization to reinteraction.  For instance, if a particle is decayed after it undergoes a hadron reinteraction, it would have to be redecayed by the secondary generator.   The workflow would also have to account for the possibility that kinematics after FSI cause the event to be thrown out by the primary generator.  Working through the several use cases should be done before making a concrete proposal for addressing these questions.  The solution could require some agreements on a standard for these items, or
additions to the common record.  

Successful implementation of this separation imposes several requirements on generator developers.  Generators would need to be configured and run in FSI-only or hard-scattering-only mode, using a common event record as input or output.  Most generators are already able to run in FSI-only or hard-scattering-only modes, but the reading and writing of records from other generators is not currently possible.  This separation also imposes requirements on the common event-record format.  Specifically, it is essential that information regarding nuclear model assumptions is recorded. This would not enforce consistency between nuclear models used in the primary and FSI stages, but would make verification of consistency straightforward.  

\subsection{Theory Interface}
\subsubsection{Motivation for a common theory interface}

The theory community generates a rich spectrum of models that attempt to describe the nucleus and how it interacts with neutrinos. To test these models, it is imperative that we compare them against experimental data. To do so most effectively, the models need to be accurately incorporated into neutrino event generators. Currently, this process is cumbersome and often requires input from generator specialists. Models also need to be individually added to each generator, requiring
additional time-consuming work.  Available theoretical models are also frequently not available in complete form, requiring extra work to adopt the model to generator requirements.  Experimental simulations require theories that can describe the full evolution from the initial state to the final state with the outgoing hadrons and leptons~\cite{ulrich}.  

These limitations lead to significant delays
between the creation of new models and their being thoroughly tested against
data. The ultimate result is an inability for current neutrino experiments to
fully benefit from recent theoretical advances in a timely manner. This
presents a worrisome challenge for future precision neutrino oscillation
measurements unless the throughput of the ``generator model pipeline" can be
significantly expanded.

It is thus desirable to define a common interface to incorporate models into generators. Ideally, this would allow theorists to provide models for testing with no need for additional work from the generator teams at the point of model generation.  The common interface would mean that experimentalists would be able to plug the model immediately into any of the available event generators (GENIE, NuWro etc), with the generator taking care of any part of the simulation process not included in the model itself.   That is, handling of the flux and detector simulation
would be delegated to the generator, as would any other aspect of event generation not considered by 
the new theory calculation.
  The interface should be simple and clearly documented, so that it does not generate significant additional burden on theorists, and so that experimentalists can quickly provide feedback on the model's compatibility with data, or with alternative models.

\subsubsection{Factors to consider for interface design}

While our goal should be to standardize the interface as much as possible, the workshop raised several issues that should be considered when evaluating possible approaches. These issues are summarized below:

\begin{enumerate}
    \item \textbf{Model scope} A wide range of models are under development, describing the nucleus and  various scattering processes, including beyond-the-Standard-Model (BSM) predictions. These theories use widely varying approaches and approximations to model the nucleus and the way it interacts with neutrinos and other particles. Some model the inclusive interaction cross section, while others concentrate on exclusive processes, such as resonant pion production, or two-particle-two-hole scattering. In this case, the generator's existing models would need to be used to supplement the rest of the inclusive cross section, and care would have to be taken to avoid the possibility of double-counting
in cases where the generator splits
up the interaction modes differently than the theory.      
    Furthermore, while some mechanisms model the full interaction process, delivering a set of final-state particle momentum vectors; others provide an alternative model only for some specific component of the interaction. For example, we may wish to test an improved model of nucleon form factors, using the generator to simulate how this affects neutrino-scattering processes. This approach would also allow easier implementation of BSM scenarios. Some of these models, for instance, include neutrino up-scattering to heavier states via new gauge bosons~\cite{Bertuzzo:2018itn, Bertuzzo:2018ftf, Ballett:2018ynz}, neutrino magnetic moment~\cite{Okun:1986na} or dark matter scattering off electrons or nuclei~\cite{deNiverville:2011it,Batell:2009di}. Each of these examples involve different physics and different theoretical motivations. Nevertheless, they can all be incorporated in neutrino generators by simply changing the leptonic tensor, which is easily calculable. In these cases, the generator will be required to simulate any unmodeled components of the interaction process, particularly final-state interactions. 
    
    The full range of models that should be supported by the proposed theory
interface has yet to be established and remains a key question that must
be answered for the effort to be successful. The desire to produce an interface
in which a new model can be inserted seamlessly into any of the standard
generators, with that software ready to simulate any unmodeled components,
makes defining an appropriate scope particularly important.

    \item \textbf{Viable output formats} Related to the previous remark, the wide range of calculations lend themselves to different ways of presenting results. For a model that simulates the full interaction chain, it may be possible to present a standardised API where a neutrino 4-momentum and target nucleus are passed to a plugin routine, which returns a vector of final-state particle PDG codes and momenta. This, however, would not be practical for models that provide only a specific component that contributes to the interaction---a leptonic or hadronic tensor, for example, or a structure function. It is understood that other models provide energy-dependent interaction cross sections, but leave the simulation of final-state particles to the generator. The simplest way to present this type of data is in the form of lookup tables, from which the relevant information can be interpolated. In this case, the probabilistic part of the simulation will take place in the generator.
    
    For some models, there is some flexibility in which information is presented. For example, a model that implements a full-chain simulation (and could thus provide individual particle momenta) may include a leptonic and/or hadronic tensor in its calculation, and could be used to generate cross section histograms or tables. This would, however, involve loss of information about the individual particle kinematics - information which may be valuable when comparing to data. 
    
    It is important that we understand the potential output information available, and ensure that a mechanism is in place to do any additional simulation needed to supplement the information provided by the model. 
    
	\item \textbf{Evaluating systematic uncertainty through reweighting} Experiments commonly use reweighting techniques to evaluate the effect of model uncertainty on their measurements. For example, when GENIE generates an event, it can later calculate a weight for the relative probability of such an event occurring if, for example, a resonant pion production parameter were increased by one standard deviation. This method is vital for evaluating systematic uncertainties on measurements and comparing models to data. Any theory interface should ensure that this functionality is maintained. In the case of a programmatic interface, this may mean additional input parameters that allow constants to be adjusted. In the case of a table- or histogram-based interface, depending on what is exposed, it could require theorists to re-run their calculations many times, raising potential time and disk-space concerns. Experiments resort to ad hoc manipulations of tables of this kind in the
absence of more rigorous theory guidance.
 Several experiments (MINERvA,
MicroBooNE, etc.) have developed weight calculators to manipulate the
table-based cross section calculation for the GENIE implementation of the
Valencia MEC model. These manipulations tend to be empirically- rather than
theory-driven.

	\item \textbf{Human factors} Our interface should be designed for ease of use, and should consider the skills and limitations of the theorists likely to use it. It was pointed out that many theorists are PhD students or postdoctoral researchers working on limited-term contracts. To fit in with this way of working, it should be possible to develop, implement, and test a model against data on timescales of the order of a year. We must also bear in mind that many theorists are not primarily programmers, and that models may be developed using tools, such as Mathematica, that are not natively compatible with the languages used in generator software. If we restricted ourselves to an interface in a particular programming language, we could severely limit the accessibility of the interface to new models.
\end{enumerate}

\subsubsection{Possible approaches}
\label{aaproaches}
One of the bottlenecks to testing new theoretical or phenomenological ideas is implementing the effect into event generators.    At the workshop, we discussed different approaches to mitigate this problem.
\begin{itemize}
  \item \textbf{Table approach}: The basic concept of the table approach is relatively simple. Theorists will be asked to provide, in a standard format, a differential cross section in some combination of variables.  For example, one currently used combination of variables is q and $\omega$, which represent the magnitude of the 3-momentum
transfer and the energy transfer to the hadronic system, respectively. The discrete information in the table is converted into a spline or interpolation model to allow for a continuous distribution. The generator will then produce a selection of events following this distribution and will be responsible for generating the actual particle kinematics for the events. 
  
GENIE employs a hadronic tensor technique to include some models 
of CCQE and CCMEC processes, including the SuSAv2~\cite{2p2hframework} and
Valencia CCMEC models~\cite{2p2hvalencia}. GENIE also includes a direct implementation of the
Valencia CCQE model that does not use table-based
 calculations.
Tables of nuclear responses, expressed as
functions of $q$ and $\omega$, are used to compute inclusive cross sections.
Both MEC implementations rely on GENIE's ``nucleon cluster model" \cite{teppei} to simulate the hadronic final state. To extend the table approach to include Ab initio computations, new 3p3h calculations and pion production models, table of responses for several nuclei are needed, as well as a recipe for how to combine the models with nuclear effects.

There are some disadvantages to the table approach: namely, a potentially large number of tables for each nucleus, neutrino scattering process, and neutrino energy. Another issue is that these tables are fully inclusive and therefore limited
in their predictive power. Any information about hadronic final states that was
provided by the model used to produce them has already been integrated out.

  \item \textbf{Hard-scatter events}:  This strategy is based on an interface developed by the Collider Physics community\cite{interfacecollider,interfacecollider2}. It relies on a factorization based on energy scales.    The factorization is correct up to power-suppressed corrections in the energy scale making the calculations more and more precise as the energy scales increase. Power-suppressed physics such as hadronization is dealt by using more phenomenological models.  Using this approach, theorists calculate events using perturbative QCD that is valid within some range of the typical energy scales involved. These events are inherently inclusive;
that is, the event has a final state of partons and possibly other non-color charged particles. The shower Monte Carlo will further branch/decay
the partons and eventually hadronize the colored final state of partons into a colorless final state of hadrons. The role of the Shower Monte Carlo is to transition the inclusive perturbative partonic calculation into a fully exclusive final state of hadrons. Current neutrino-nucleon scattering experiments are conducted at relative low energy scales.  Only in the DIS regime one can directly port the approach described above.
At lower energies no clear factorization between hard scattering and subsequent decay is possible. Issues such as nuclear effects and final state interactions are important, if not dominant.
Still one can use the methodology based on factorization used for the Collider approach as a phenomenological model to build more flexible generators for neutrino physics. There is the core scattering and subsequent further decay/showering using FSI until a stable hadronic final state emerges.
  \item \textbf{Interface using lepton and/or hadronic tensors} $L_{\mu\nu}$ and $W_{\mu\nu}$. Theorists would provide one or both tensors, precalculated and presented in a standard format or provide code that computes these.  Simulation would take place within the generator, using these values for the tensors.
  \item \textbf{Fortran interface}: An alternative approach is to develop a uniform computer code format that allows the theorist to implement cross section calculations directly into the event generator.   The challenges with this approach are that it requires a choice of computing language and will likely be limited to calculations where the Monte Carlo integration is relatively simple. An example would be a Fortran
wrapper that attaches an event generator (typically written in C++) to theory code (typically written in Fortran) that computes structure functions, form
factors, nuclear responses, or differential cross sections.
 An example was presented at the workshop, a Fortran wrapper that takes the structure functions F1, F2, and F3, using the deep inelastic interaction calculation from H. Haider \cite{huma2}. The procedure was validated by comparing the output of GENIE, output of the theory model, and data from Neutrinos At the Tevatron and CERN--Dormund--Heidelberg--Saclay--Warsaw (page 13 from \cite{steven}). 

  \end{itemize}
\subsubsection{New accurate models available }
Some of the available models in current neutrino event generators are based on old calculations performed in the 1980s.  A notable exception is GIBUU, which is the only event generator that is based on quantum-kinetic transport, the state-of-the-art method to describe incoherent nuclear reactions. It is also the only theoretical framework for a description of neutrino-nucleus reactions that strives for consistency between the various subprocesses. 

New approaches with more rigorous calculations are becoming available. At the workshop we discussed two modern calculations: the spectral function, and the short-time approximation. The key additions for these modern calculations are:
\begin{itemize}
  \item The Spectral Function framework is based on the impulse approximation and realistic spectral functions, which are two dimensional distributions linking nucleon momentum and binding energy. This scheme combines a realistic description of the initial target state with a fully relativistic interaction vertex and kinematics. The results obtained for the electron--nucleus cross-section in the QE region show that this method is in remarkably good agreement with data. The factorization scheme has been extended and generalized to include two-nucleon emission processes induced by relativistic MECs and pion-production mechanisms \cite{noemi,noemi2,noemi3}. 

\item A novel approach called the short-time approximation, based on realistic models of nuclear interactions and currents, is becoming available to evaluate the inclusive and exclusive responses of nuclei. The approach accounts reliably for crucial two-nucleon dynamics, including correlations and currents, and provides information on back-to-back nucleons observed in electron and neutrino experiments \cite{saori}.  
For the first time, the interferences between 1p1h and 2p2h are available.  
\end{itemize}

A survey was carried out to collect details about available models that are not currently implemented in neutrino event generators. Table \ref{table_models} shows the results of this survey. One question from the survey asked theorists to specify if their calculations could provide the leptonic and hadronic tensors $L_{\mu\nu}$ and $W_{\mu\nu}$; eleven of the models could provide this information. The leptonic and hadronic tensors are therefore likely to be inputs from the theory community for many models, and an interface will be written to incorporate the models into event generators.  A full report of the survey  is available at \cite{survey}.
\begin{table}[ht]

\centering

\caption{Summary of responders to the neutrino interaction modeling survey}
\vspace{2mm}
\begin{tabular}{|L|L|}
\hline
Authors&Processes \\
\hline
Saori Pastore et al. \cite{saori}&QE and MEC\\
\hline
Gil Paz et al.\cite{Gil}&QE\\
\hline
Artur Ankowski et al.\cite{Artur}&QE\\
\hline
Alessandro Lovato et al.\cite{Alessandro}-\cite{Alessandro4}&Elastic scattering, low energy transition, QE\\
\hline
Luis Alvarez et al.\cite{Luis}-\cite{Luis7}&QE, (coherent) pion, eta production and photon emission\\
\hline
Noemi Rocco et al.\cite{noemi}-\cite{noemi3}&QE, MEC, 1 and 2 pion production\\
\hline
Raul Jimenez et al.\cite{Raul}-\cite{Raul2}&QE\\
\hline
Minoo Kabirnezhad. et al.\cite{Minoo}-\cite{Minoo2}&Single pion production\\
\hline
Natalie Jachowicz et al.  \cite{Natalie}-\cite{Natalie9}& Elastic scattering, low-energy excitations, QE, MEC, SRC and single pion production\\
\hline
Toru Sato et al.\cite{Toru}&Meson(pion,kaon,eta,2pi) production for nucleon in nucleon resonance region\\
\hline
Huma Haider et al.\cite{Huma}&Deep Inelastic Scattering \\
\hline
Juan Nieves et al. \cite{Juan}-\cite{Juan26}&QE+SpectralFunctions+RPA+2p2h+pion production (Delta, chiral background, some other N*) \\
\hline
Maria Barbaro et. al.
\cite{Maria}&Quasi-elastic scattering, two-nucleon emission (2p2h), pion production, higher resonances, deep inelastic scattering, both CC and NC processes.\\
\hline
\end{tabular}
\label{table_models}
\end{table}

\section{Work Plan}
\label{sec:plan}

\subsection{Flux/Geometry Driver and Formats}

A working group is now being formed, which is tasked with pursuing prototypes of both of the common flux driver options outlined in section~\ref{sec:flux_summary}, considering the advantages and challenges of each option, choosing one option and implementing it.  Hugh Gallagher, Yoshinari Hayato, and Luke Pickering have agreed to co-lead the group.    The initial goal of the group will be to have a common flux driver developed, documented and implemented in at least one new generator by summer 2021.   

One path that may be investigated by this working group is new functionality developed by Chris Backhouse (UCL) for NOvA that allows the GENIE event generator to draw from a library of events generated by an external generator, thus making use of the GENIE geometry driver and providing events in a GENIE format for consumption by experiment simulation chains.  Tools exist to create libraries from GiBUU, and it is expected that creating similar tools for NuWro and NEUT would be straightforward.  

\subsection{Output Format and Separation of Primary Interaction and Final State Interactions}

\begin{figure} [h] 
\begin{center}
\includegraphics[width=0.7\textwidth]{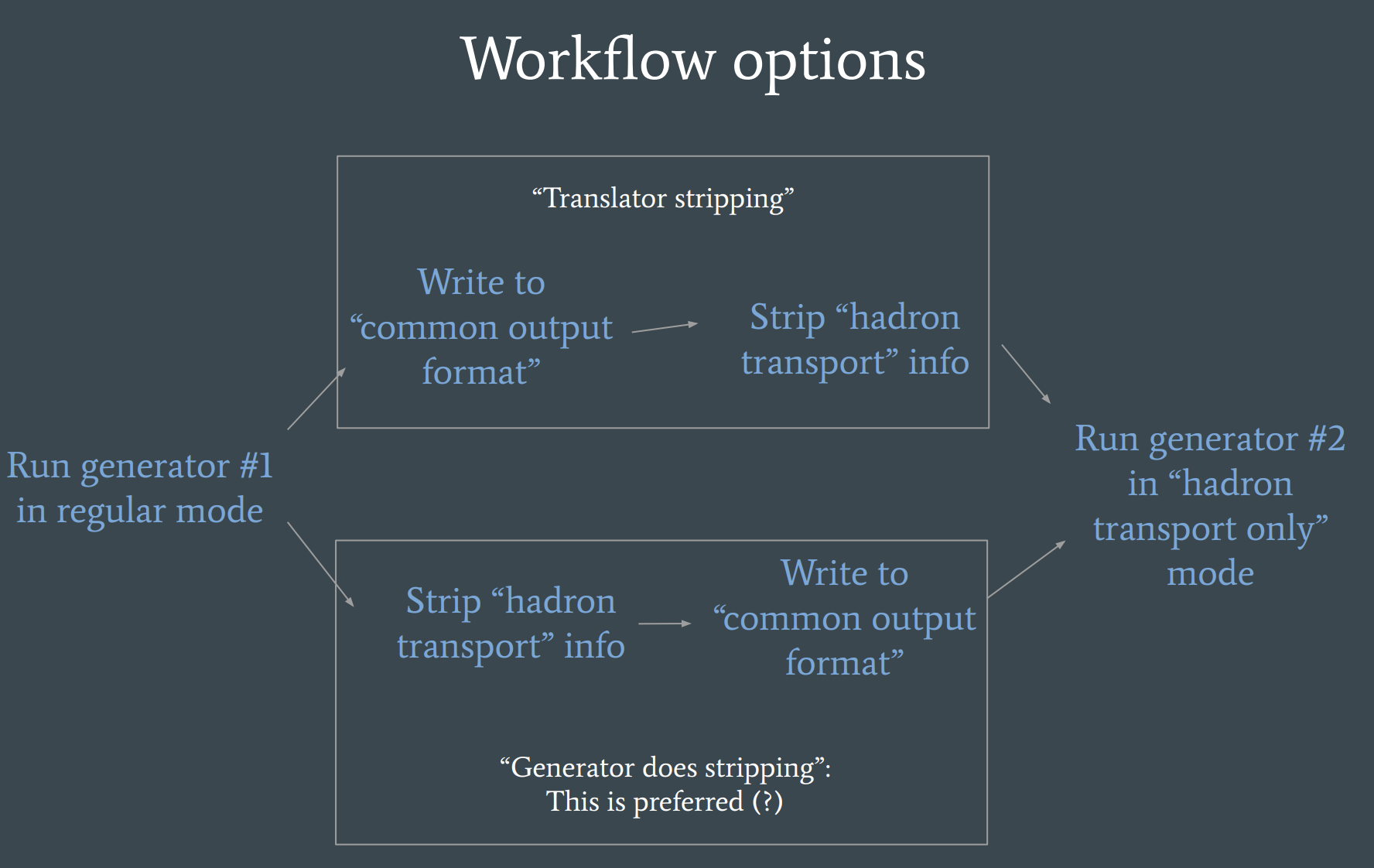}
\end{center}
\caption{\label{fig:fsi}Schematic of possible hard-scattering/FSI separation workflows.  Figure courtesy Jeremy Wolcott.   }
\end{figure}

A working group is being formed that will build a proposal for a common event format.  The group will include:
\begin{itemize}
    \item Event generator representatives
    \item Experts on existing storage formats including HepMC
    \item At least one representative of LArSoft
    \item At least one representative of NUISANCE, which has experience forming a basic common event format  
    \item Representatives from current experiment end users who are fitting/tuning/comparing different models to data
    \end{itemize}
Josh Isaacson and Clarence Wret have agreed to lead this group.  Once the group has identified a proposal, they will review it with participants of the workshop and other interested parties, and iterate as necessary.  

This working group will also be responsible for developing at least one example translator between an existing generator format and the new format.  The group will also initially develop at least one representative consumer of the common format.  For example, NUISANCE could be migrated to use the new format, or a generator could be modified to use this format as part of developing factorized FSI.  The group will be responsible for testing these implementations and reporting on results and lessons learned.  

Since the factorization of the primary interaction and FSI is heavily linked to the common event format, we propose that the event format group also pursue work regarding the factorization.  The initial task here would be to develop a ``first stage" (primary neutrino interaction) in one generator and a ``second stage" (FSI) in another generator.  The first stage could be done in a translator if necessary, stripping off any information about FSI in the first generator, although there is a desire to have the stripping done in the initial generator itself (see workflow options shown in Fig.~\ref{fig:fsi}).  The GiBUU generator already contains functionality to read in information from the primary interaction and run its hadron transport, so it is a good candidate for the initial ``second stage" generator.

\subsection{Theory Interface}
A working group comprising neutrino theorists, collider theorists, and neutrino experimentalists has been in existence since 2017 at FNAL. The group identified the crucial modeling needed for DUNE and has been working to incorporate accurate models in the event generator used by DUNE.  Members of the group are Fermilab staff, fellows, and distinguished scholars, as well as colleagues from universities and other laboratories. Several post-docs and graduate students participate. GENIE developers play a key role. One of the goals is to create an interface between theory and generators such as GENIE and work together toward improving the models and incorporating them into the simulations. Some of the topics under study are ab initio nuclear models, pion production, deep inelastic scattering in the nucleus at a few GeV, lattice-QCD calculations of form factors, radiative corrections, and relating electron and neutrino cross sections.

The group meets once a month to discuss progress and new ideas. Slides are posted at \url{https://indico.fnal.gov/category/724/}.  They are in the progress of producing first results for some of the contributions, such as the implementation of two calculations in the event generator GENIE.

Currently, the group has a few graduate students and a small fraction of a postdoc carrying out the calculations and implementing the models in the event generator. The progress has been slow; more resources are needed to fully implement the possible approaches outlined in section~\ref{aaproaches}.

The development of the theory interface will be led by the working group at FNAL. Currently there are three conveners; Minerba Betancourt, Adi Ashkenazi and Walter Giele.

The first step will be to write an interface that reads the leptonic  $L_{\mu\nu}$ and hadronic $W_{\mu\nu}$ tensors from the different calculations in table \ref{table_models}. The following steps will be to write a Fortran interface, and an interface to read hard scatter events. 
  
\subsection{Current Commitments to these activities}
\begin{itemize}
\item Luke Pickering, Clarence Wret, Steve Dytman, and Hugh Gallagher from US DOE supported University program.
\item Robert Hatcher, Steve Mrenna, Joshua Isaacson, and Steve Gardiner from FNAL.
\item Yoshinari Hayato from ICCR/IPMU in Japan. 
\item Linda Cremonisi from QMUL
\item Jan Sobczyk from  University of Wrocław
\item  FNAL+ICCR/IPMU US-Japan proposal for 2020-2021 workshops for NEUT interface.
\end{itemize}

\end{document}